\documentclass[a4paper,11pt]{article}
\usepackage{pos}
\usepackage{wrapfig}
\usepackage[export]{adjustbox}
\usepackage{xargs}
\usepackage[colorinlistoftodos,prependcaption,textsize=footnotesize]{todonotes}
\newcommandx{\DM}[2][1=]{\todo[linecolor=orange,backgroundcolor=orange!25,bordercolor=orange,#1]{DM: #2}}
\newcommandx{\KH}[2][1=]{\todo[linecolor=green,backgroundcolor=green!25,bordercolor=green,#1]{KH: #2}}
\newcommandx{\UW}[2][1=]{\todo[linecolor=blue,backgroundcolor=blue!25,bordercolor=blue,#1]{UW: #2}}
\newcommandx{\AI}[2][1=]{\todo[linecolor=red,backgroundcolor=red!25,bordercolor=red,#1]{AI: #2}}

\title{Fixed point actions from convolutional neural networks}
\ShortTitle{FP actions from L-CNNs}

\author[a]{Kieran Holland}
\author[b]{Andreas Ipp}
\author[b]{David I.~M\"{u}ller}
\author*[c]{Urs Wenger}

\affiliation[a]{University of the Pacific,
3601 Pacific Ave., Stockton, CA 95211, USA}
\affiliation[b]{Institute for Theoretical Physics, TU Wien, Wiedner
  Hauptstraße  8-10/136, A-1040 Vienna, Austria}
\affiliation[c]{Albert Einstein Center for Fundamental Physics,
Institute for Theoretical Physics, University of Bern, Sidlerstraße 5, 3012 Bern, Switzerland}

\emailAdd{kholland@pacific.edu}
\emailAdd{ipp@hep.itp.tuwien.ac.at}
\emailAdd{dmueller@hep.itp.tuwien.ac.at}
\emailAdd{wenger@itp.unibe.ch}

\abstract{Lattice gauge-equivariant convolutional neural networks (L-CNNs) can be used to form arbitrarily shaped Wilson loops and can approximate any gauge-covariant or gauge-invariant function on the lattice. Here we use L-CNNs to describe fixed point (FP) actions which are based on renormalization group transformations. FP actions are classically perfect, i.e., they have no lattice artifacts on classical gauge-field configurations satisfying the equations of motion, and therefore possess scale invariant instanton solutions. FP actions are tree–level Symanzik–improved to all orders in the lattice spacing and can produce physical predictions with very small lattice artifacts even on coarse lattices. We find that L-CNNs are much more accurate at parametrizing the FP action compared to older approaches. They may therefore provide a way to circumvent critical slowing down and topological freezing towards the continuum limit. }

\FullConference{The 40th International Symposium on Lattice Field Theory (Lattice 2023)\\
July 31st - August 4th, 2023\\
Fermi National Accelerator Laboratory\\}

\begin{document}
\maketitle

\section{Introduction}
\begin{figure}
  \includegraphics[width=0.28\textwidth]{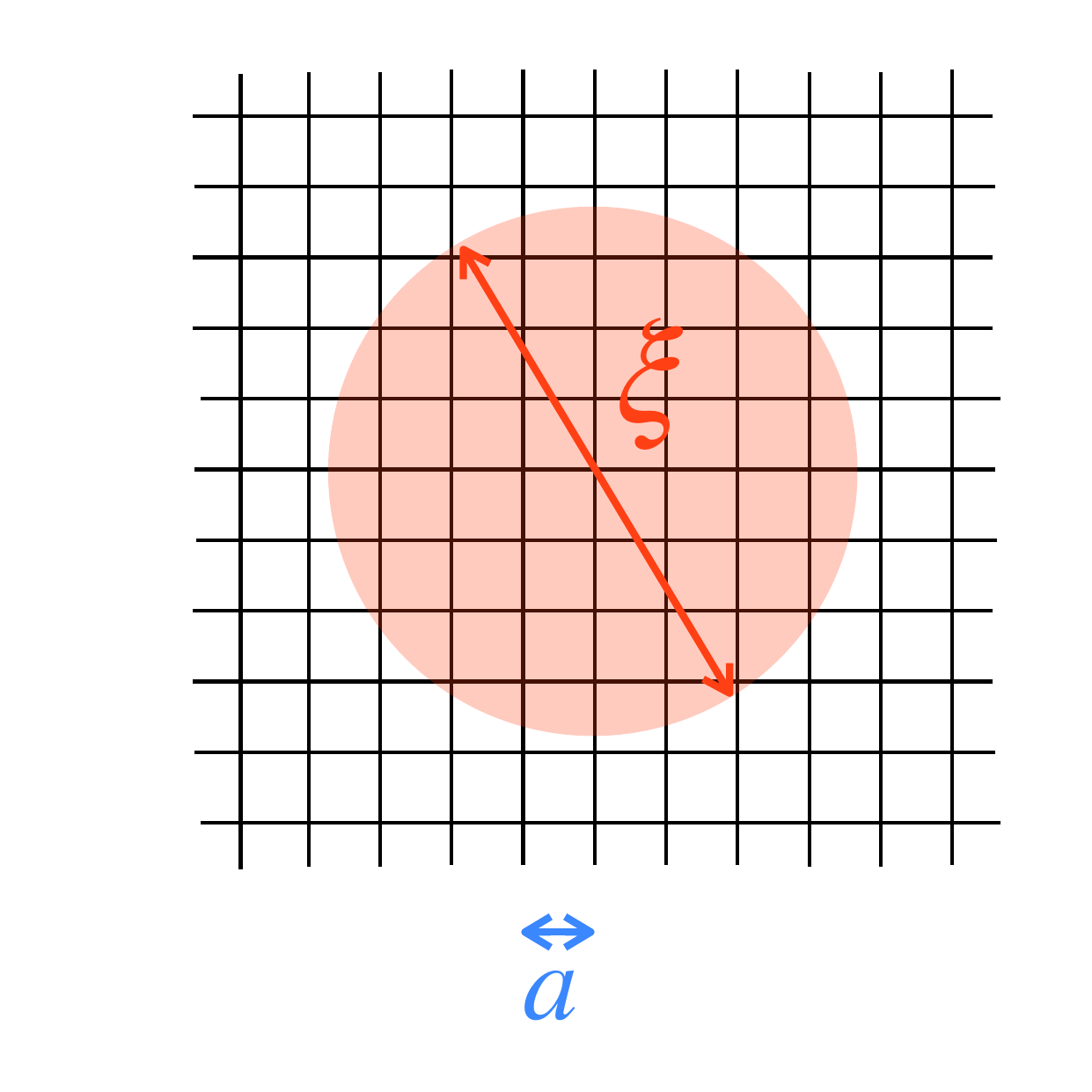}
  \hfill
 \includegraphics[width=0.28\textwidth]{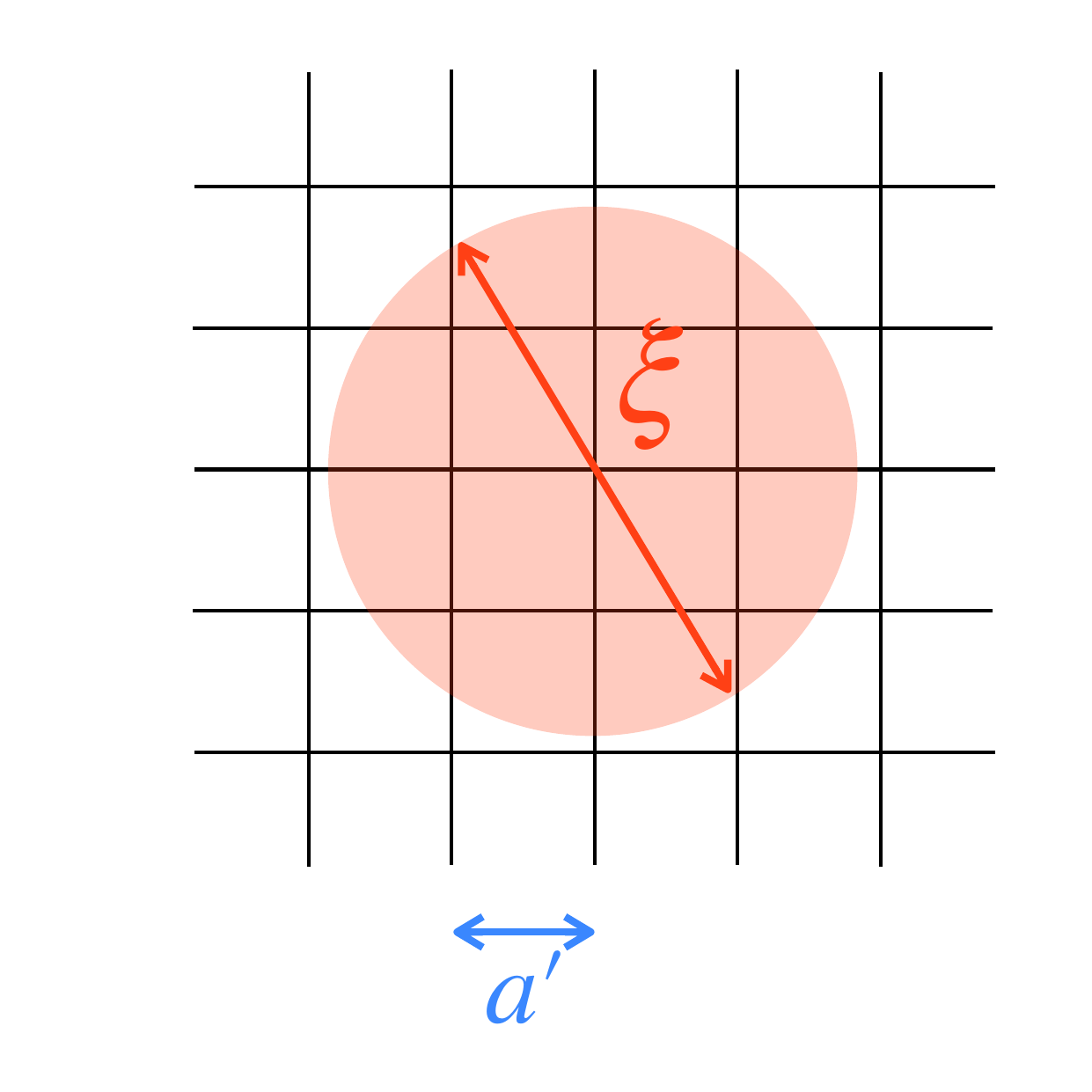}
  \hfill
  \includegraphics[width=0.28\textwidth]{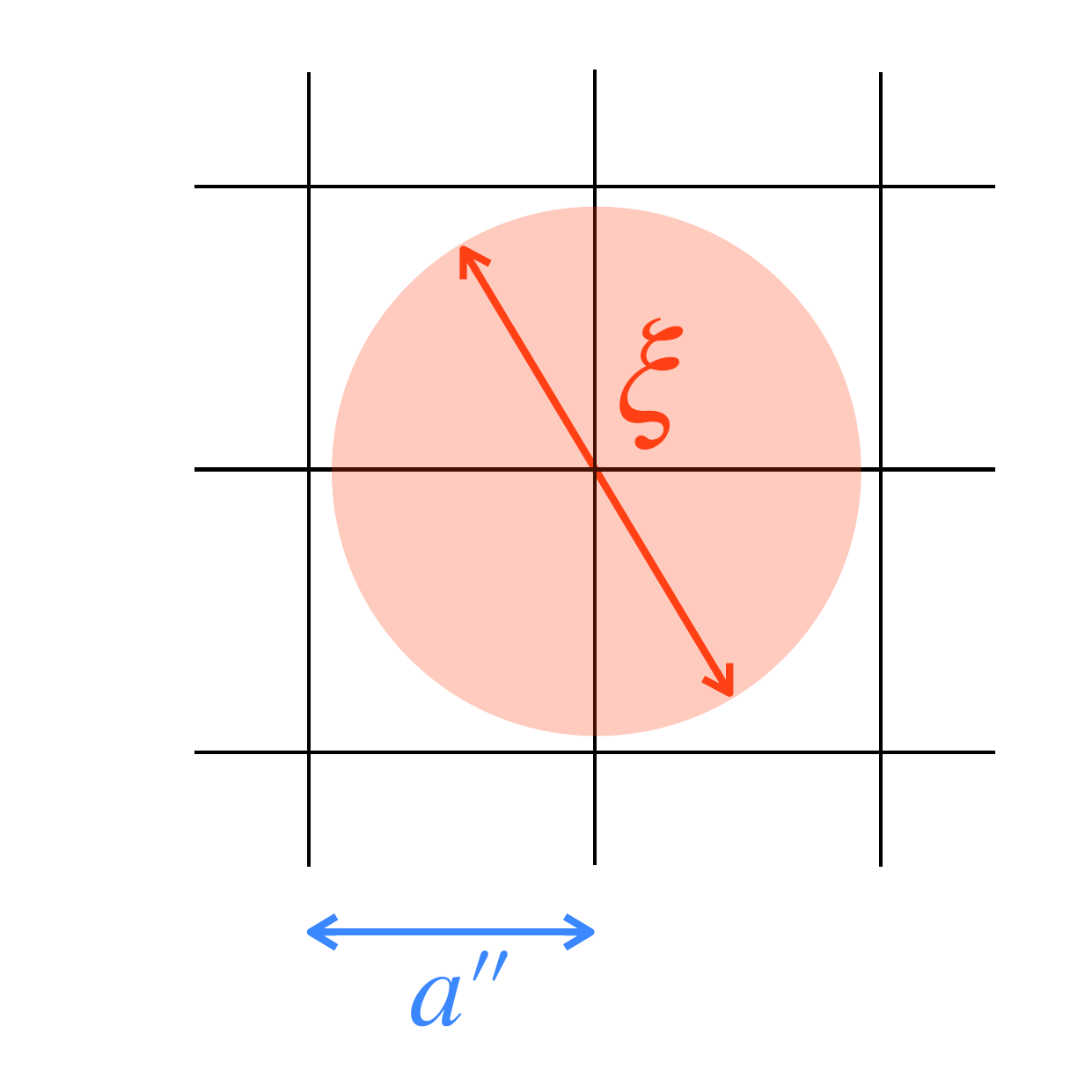}
   \caption{Sketch of the procedure of taking the continuum limit: as
    the gauge coupling is decreased from right to left, $g <
    g^{'} < g^{''}$, or equivalently $\beta > \beta^{'} > \beta^{''}$ is
    increased, the lattice spacing decreases, $a < a^{'} < a^{''}$. In
    the limit  $\beta \rightarrow \infty$ the lattice spacing $a\rightarrow
0$, i.e., $\xi/a \rightarrow \infty$, and the continuum limit is reached. Renormalization group transformations map the system from the left to the right side.
\label{fig:continuum limit}}
\end{figure}

Consider an asymptotically free gauge field theory on the lattice,
e.g., SU($N_c$) lattice gauge theory, which is described by the
partition function 
\[
Z(\beta)=\int{\cal D}U \exp\{-\beta A[U]\}
\]
with gauge coupling $\beta = {2N_c}/{g^2}$ and the gauge action
$A[U]$ which is a function of the gauge links $U$. The integration
over the gauge links is defined via the Haar measure ${\cal D}U$ of
the gauge group. Expectation values
for observables ${\cal O}_\xi[U]$ with a characteristic length scale $\xi$ are defined as
\[
\langle{\cal O}_{\xi}(\beta)\rangle=\frac{1}{Z}\int{\cal D}U
\exp\{-\beta A[U]\} \, {\cal O}_\xi[U] .
\]
A typical observable could be, for
example, a correlation function of operators whose exponential decay at
asymptotically large time separations is governed by $\xi$. When the
physical scale is expressed in units of the lattice
spacing $a$, the result $\xi/a$ is dimensionless. The lattice spacing
itself is determined by the gauge coupling, i.e., $a=a(\beta)$. Specifically, the continuum limit of the theory is
reached by taking $\beta \rightarrow \infty$, such that $a\rightarrow
0$ and $\xi/a \rightarrow \infty$. This situation is sketched in Figure
\ref{fig:continuum limit}.
From the point of view of a statistical lattice system the
continuum limit of the lattice field theory corresponds to approaching the critical point  of a
continuous (second order) phase transition where the correlation
length $\xi$ diverges. The universality at the critical point
guarantees the independence of the so-obtained physical observables
from the details of the microscopic lattice definition of the theory,
i.e., different discretizations lead to the same universal
results. This has long been understood in the context of 
renormalization group transformations (RGT). Performing a (real space)
RGT by blocking the degrees of freedom increases the lattice spacing
$a \rightarrow a' \rightarrow a''$ and maps the system from left to right in Figure \ref{fig:continuum limit},
while keeping the physical length scales unchanged. It is therefore in
principle possible to extract continuum physics, i.e., values of the 
observables in the continuum, from
systems defined at finite lattice spacings as long as the corresponding correlation length is still well defined, i.e., $\xi/a
\gtrsim 1$.

It is well known that taking the continuum limit in practice is computationally
very demanding due to the problem of {\it critical slowing down} when
approaching a critical point. Moreover, for SU($N_c$) gauge theories
{\it topological freezing} poses an additional problem \cite{Schaefer:2010hu}. It is
therefore prohibitive for lattice simulations to reach very fine
lattice spacings. On the other hand, simulations at coarse lattice
spacings are computationally cheap, but (with the usual
discretizations) not very helpful as the lattice artifacts are large
and not well controlled. This is where the RGT comes to the rescue by providing
{\it discretizations without lattice artifacts} which allow cheap
simulations at coarse lattice spacings, thereby avoiding critical
slowing down and topological freezing, as well as uncontrollable lattice artifacts at the same time (see Ref.~\cite{Hasenfratz:1998bb} for a pedagogical introduction to the topic).  

\section{Renormalization group transformations}
A real space RGT can be defined by 
averaging (blocking) the degrees of freedom on the fine lattice before integrating them out. More specifically, one has
\begin{equation}
  \exp\left\{-\beta' A'[V]\right\} = \int{\cal D}U \exp\left\{-\beta
    \left(A[U] + T[U,V]\right)\right\},
  \label{eq:RGT}
\end{equation}
where $T[U,V]$ is a blocking kernel relating the fine gauge links $U$
to the coarse gauge links $V$. For gauge theories, the blocking kernel
can be defined as
\[
T[U,V] = - \kappa \sum_{x_B, \mu} \left\{ \text{Re} \text{Tr} \left(V_\mu(x_B) \cdot Q^\dagger_\mu(x_B) \right) - {\cal N}_\mu^\beta\right\},
\]
where the sum is over the lattice sites $x_B$ of the coarse lattice
and the normalization factor ${\cal N}_\mu^\beta$ guarantees
$Z(\beta') = Z(\beta)$, i.e., unchanged long distance physics. The
blocked link $Q_\mu(x_B)$ is obtained by first smearing the fine links
using a linear combination of the original link with planar, spatially
diagonal and hyperdiagonal staples according to
\[
S_\mu^\text{smeared} = s_0 \cdot
\includegraphics[valign=m,width=2cm]{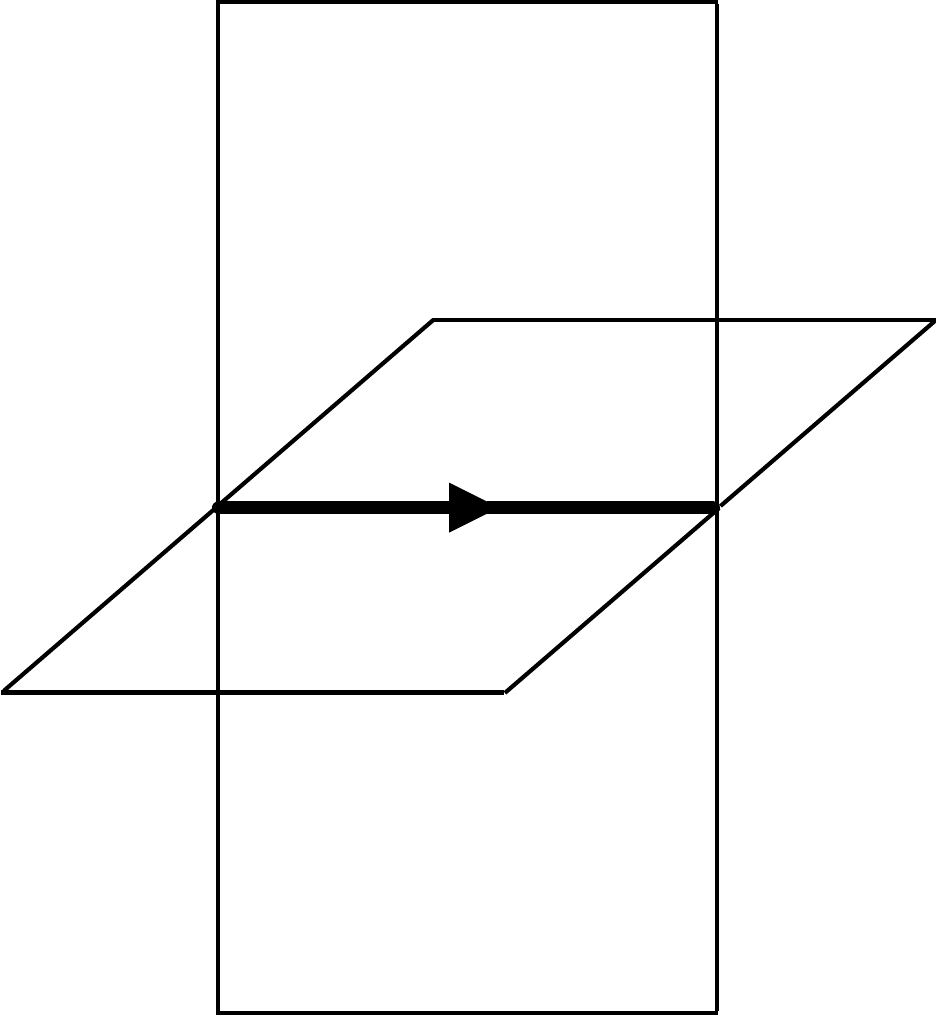} +  s_\text{pl}
\cdot \includegraphics[valign=m,width=2cm]{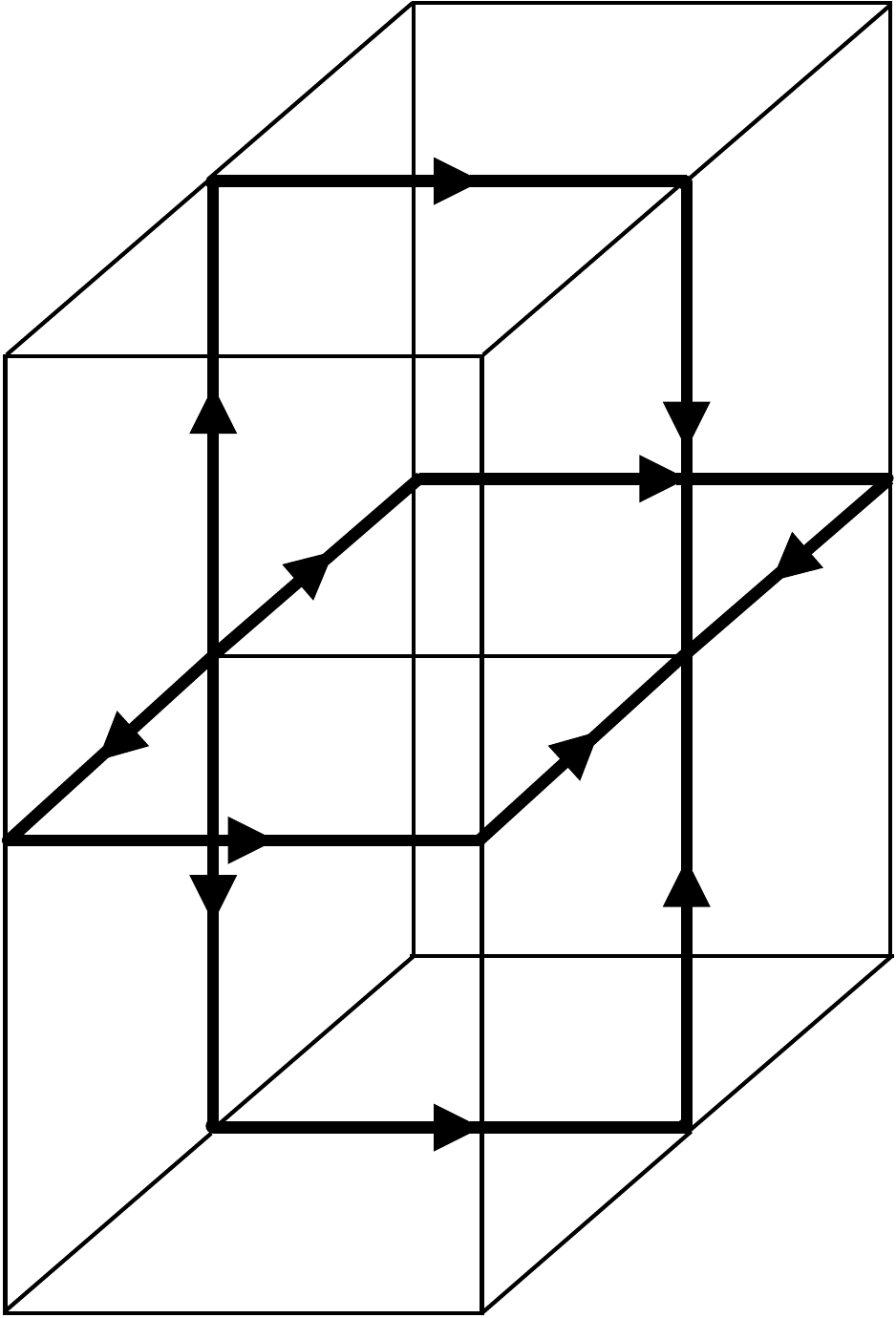} + s_\text{d}
\cdot \includegraphics[valign=m,width=2cm]{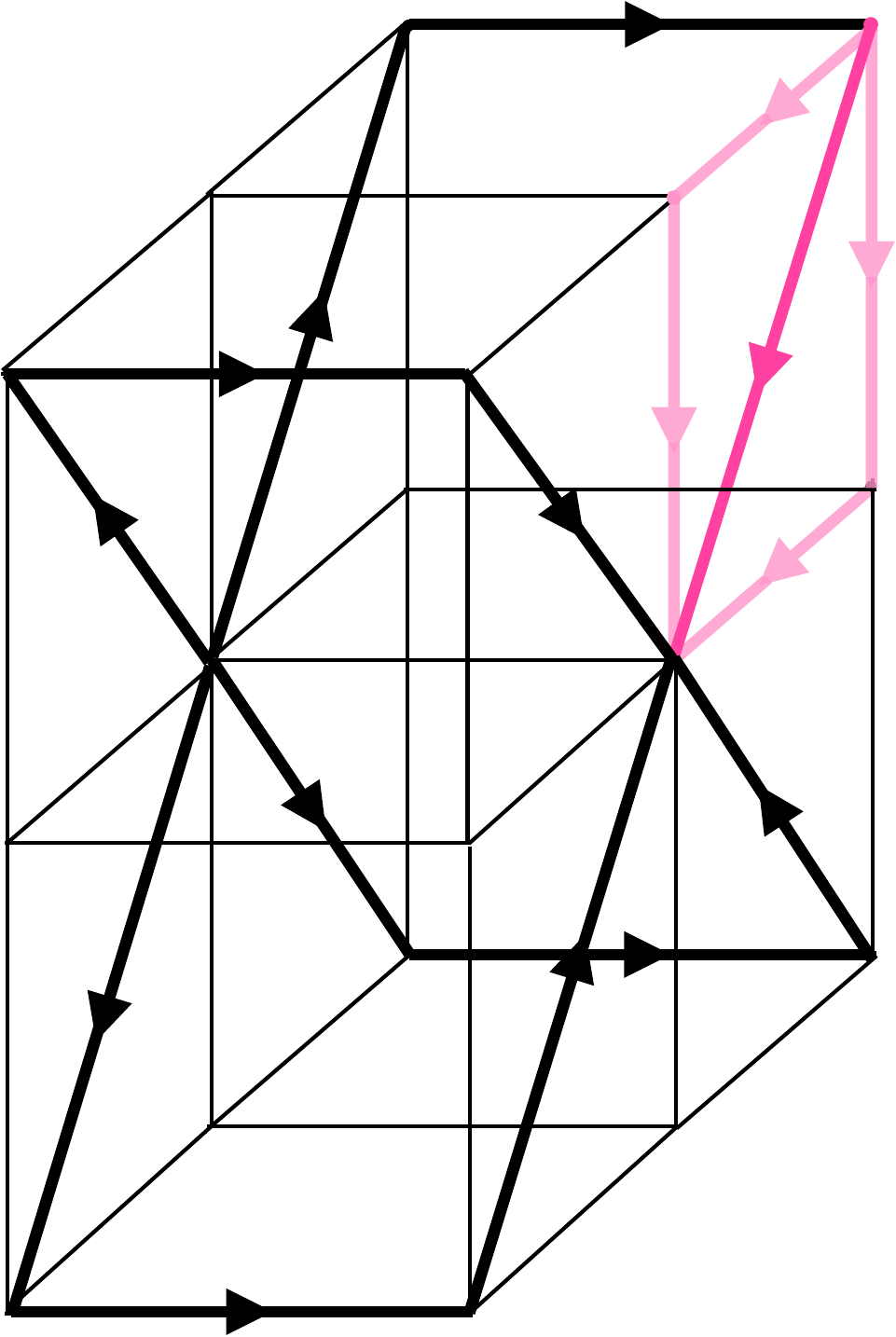} + s_\text{hd}
\cdot \ldots,
\]
with $s_0, s_\text{pl}, s_\text{d}, s_\text{hd}$ being arbitrary constants \cite{Blatter:1996np}. Note
that the smeared links $S_\mu^\text{smeared}$ are no longer an element
of the gauge group. These smeared links are now multiplied together
such that they connect lattice sites corresponding to the ones on the
coarse lattice,
\[
Q_\mu(x_B) = S_\mu^\text{smeared}(x) \cdot  S_\mu^\text{smeared}(x+\hat\mu)
\, . 
\]
It is easy to see that this procedure produces a linear combination of
a plethora of gauge link paths connecting $x$ and $x+2\hat \mu$ and taking all
links within the attached hypercubes into account.  

\begin{figure}
  \centering
  \includegraphics[width=0.6\textwidth]{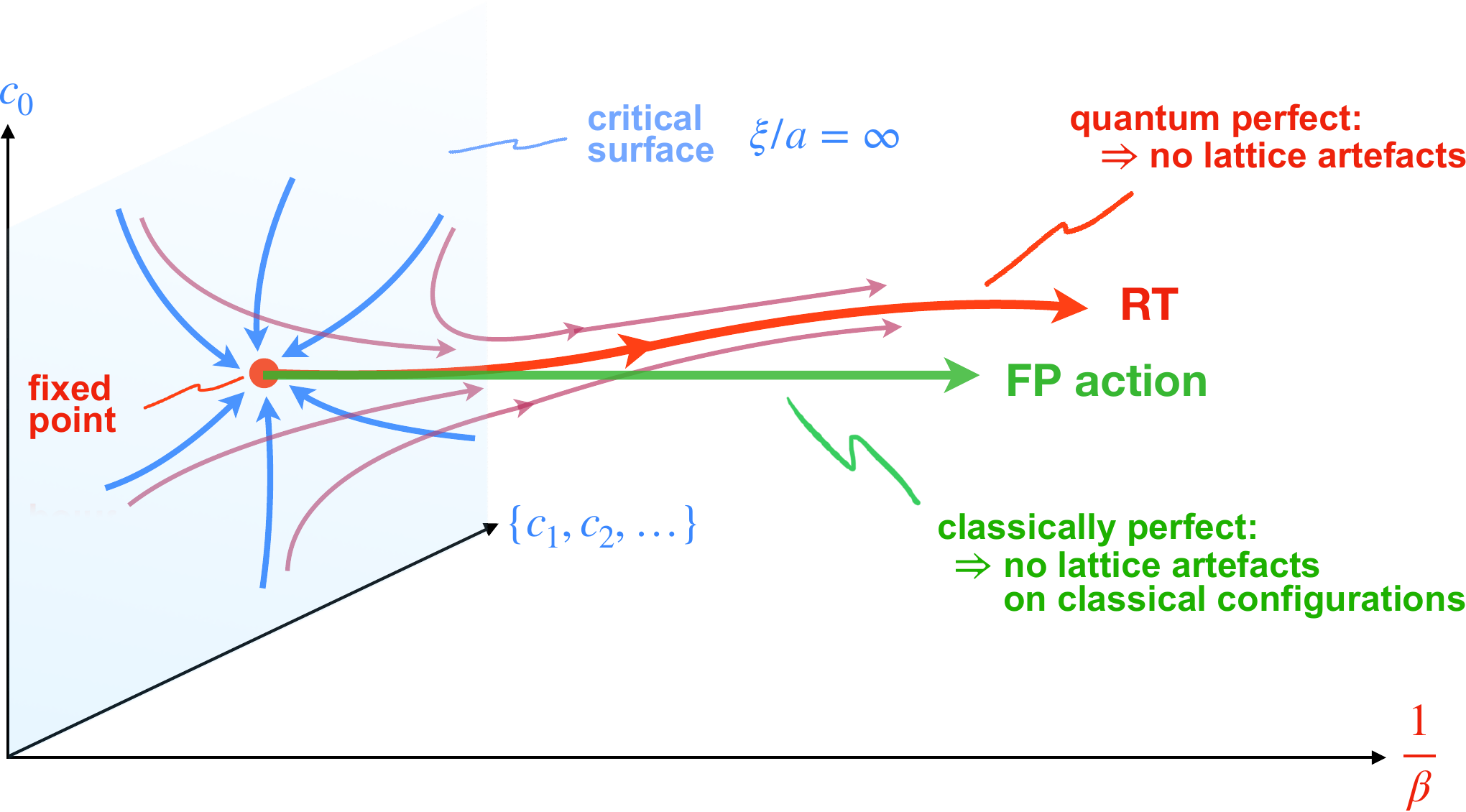}
  \caption{Illustration of the RGT flow in the infinite-dimensional
    space of couplings. Under repeated RGTs the couplings approach
    the renormalized trajectory (RT), unless one starts on the
    critical surface in which case the couplings flow into the FP of
    the RGT. Figure taken from \cite{Holland:2023aaa}. \label{fig:RGT}}
\end{figure}

The effective action $\beta' A'[V]$ on the LHS of Eq.~(\ref{eq:RGT}) is in general described by infinitely many couplings
$\{c'_\alpha\}$. Repeating the RGT yields a sequence of sets of
couplings which maps out a flow in the
infinite dimensional coupling space as illustrated in Figure
\ref{fig:RGT}.
Under repeated RGTs the couplings approach the renormalized trajectory
(RT), unless the RGT procedure starts from a set of couplings on the
critical surface where $\xi/a=\infty$. In that case the couplings flow
into the fixed point (FP) of the RGT defined by
$\{c^\text{FP}_\alpha\} \stackrel{\text{RGT}}{\longrightarrow}
\{c^\text{FP}_\alpha\}$. Note that the infrared physics described
by the actions on the RT is unchanged since the RGT only integrates out ultraviolet modes.
 Hence, actions defined along the RT reproduce continuum
physics without any lattice artifacts and are therefore {\it quantum
  perfect}.

Finding such quantum perfect actions faces two practical
challenges, namely {\it (a)} how to parametrize actions on the RT,
i.e., how to choose a necessarily finite set of couplings $\{c_\alpha\}$, and
{\it (b)} how to determine $\{c^\text{RT}_\alpha\}$ or
$\{c^\text{FP}_\alpha\}$. For asymptotically free theories, the latter
problem has been solved by Hasenfratz and Niedermayer in Ref.~\cite{Hasenfratz:1993sp}.
They observed that in the limit $\beta\rightarrow\infty$ (on the
critical surface) the integration on the RHS of Eq.~(\ref{eq:RGT})
is dominated by the minimizing configuration 
and hence
becomes a classical saddle point problem independent of $\beta$,
\begin{equation}
  A^\text{FP}[V] = \min_{\{U\}} \left\{ A^\text{FP}[U] +
    T[U,V]\right\} \, .
\label{eq:FP equation}
\end{equation}
They subsequently  showed  that the corresponding action employed at finite values of $\beta$
along the straight line emanating from the FP on the critical surface
is {\it classically perfect}. This means that there are no discretization effects
 when the action is evaluated on configurations
fulfilling the classical equations of motions. As a
consequence, lattice artifacts of ${\cal O}(a^{2n})$ are absent for all
$n$. While the FP action is not quantum perfect, i.e., lattice artifacts
of ${\cal O}(g^2 a^{2n})$ are present, it is expected that these
effects are suppressed sufficiently close to the critical surface
where the couplings of the FP action closely follow the RT. That this
is indeed the case has been demonstrated in Ref.~\cite{Niedermayer:2000yx} where a rich parametrization of the FP
action was investigated in Monte Carlo simulations based on the RGT defined in Ref.~\cite{Blatter:1996np}. The results for the deconfinement phase
transition, the static quark-antiquark potential, and the glueball
mass spectrum showed only very small lattice artifacts, if any, up to
lattice spacings as coarse as $a \sim 0.33$ fm.

\section{Machine learning the FP action}
\begin{figure}
  \centering
  \includegraphics[width=0.83\textwidth]{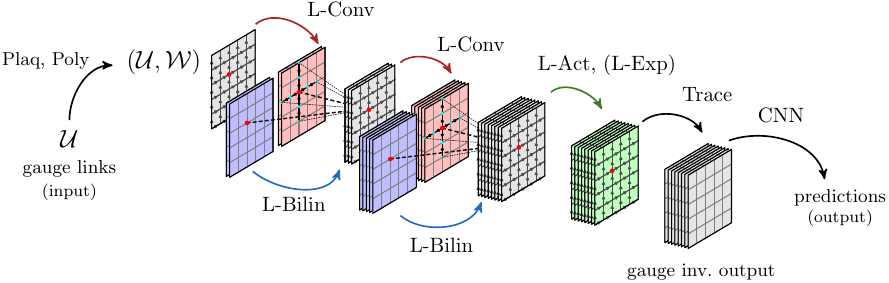}
  \caption{Illustration of a particular lattice gauge-equivariant convolutional neural
    network (L-CNN) architecture, cf.~text and Ref.~\cite{Favoni:2020reg} for further details. \label{fig:L-CNN architecture}}
\end{figure}

The second challenge in the construction of FP actions is to find a
suitable parametrization. Here we make use of the recent developments
in machine learning (ML) architectures. In Refs.~\cite{Favoni:2020reg, Aronsson:2023rli} a lattice gauge-equivariant convolutional neural
network (L-CNN) was constructed which is capable of learning any gauge-covariant or gauge-invariant function of gauge fields on a lattice. As
such, the architecture is predestined to accurately describe FP
actions. The key elements of the L-CNN architecture are the
convolutional (L-Conv) and the bilinear (L-Bilin) layers. The L-Conv layer parallel-transports gauge-covariant
objects $W$ from a finite region around the lattice site $x$ (the \emph{receptive field}) according to
\[
  W^\text{L-Conv}_{i}(x) = \sum_{j,\mu,k} \omega_{i,j,\mu,k} U_{k\cdot \mu}(x) W_j(x+k\cdot \hat \mu) U^\dagger_{k\cdot \mu}(x),
\]
where $U_{k\cdot \mu}(x)$ is the product of gauge links along a path
connecting  $x$ to $x+k\cdot \hat \mu$. The receptive field is determined by the kernel size $K$ with $|k| < K$. The indices $i,j$ label the channels of the data and $\omega$ are trainable parameters of the layer.
The L-Bilin layer produces new
gauge-covariant objects by forming bilinear combinations of two gauge-covariant objects $W$ and $W'$ at  lattice site $x$,
\[
W^\text{L-Bilin}_i(x) = \sum_{j,j'}\alpha_{i,j,j'} W_j(x)
W'_{j'}(x) \, ,
\]
where $\alpha$ are trainable parameters.
Additional layers contain activation functions (L-Act), exponentiation
(L-Exp), or tracing (Trace) of the gauge-covariant objects at each
lattice site. 
The input of the L-CNN is a gauge link configuration and the plaquettes at every lattice site.
An example of the full architecture is sketched in Figure
\ref{fig:L-CNN architecture}.
With this architecture, it is possible to recursively generate
combinations of arbitrarily complicated closed loops of gauge links of
any shape. Hence, any contribution to the FP action can in principle
be generated and described by the L-CNN.

The data for learning the FP action is generated as follows. For a given
coarse gauge field configuration $V$ the FP action value is determined
by the sequence of minimizing configurations $U, U',\ldots$ according
to an inception procedure defined by iterating the FP Eq.~(\ref{eq:FP equation}),
\[
  A^\text{FP}[V] = \min_{\{U\}} \left\{ A^\text{FP}[U] +
    T[U,V]\right\} =\min_{\{U',U\}}\left\{A^\text{FP}[U'] + T[U',U]
    +T[U,V]\right\} = \ldots\, .
\]
The so-obtained exact FP action values $A^\text{FP}[V]$ are used for training, testing, and validation. In
practice, we only perform one iteration of the procedure and use an existing,
sufficiently good parametrization of the FP action on the RHS of Eq.~(\ref{eq:FP equation}) such that the error on $ A^\text{FP}[V]$ is well controlled.

\begin{figure}[t!]
\center
   \includegraphics[width=0.45\textwidth]{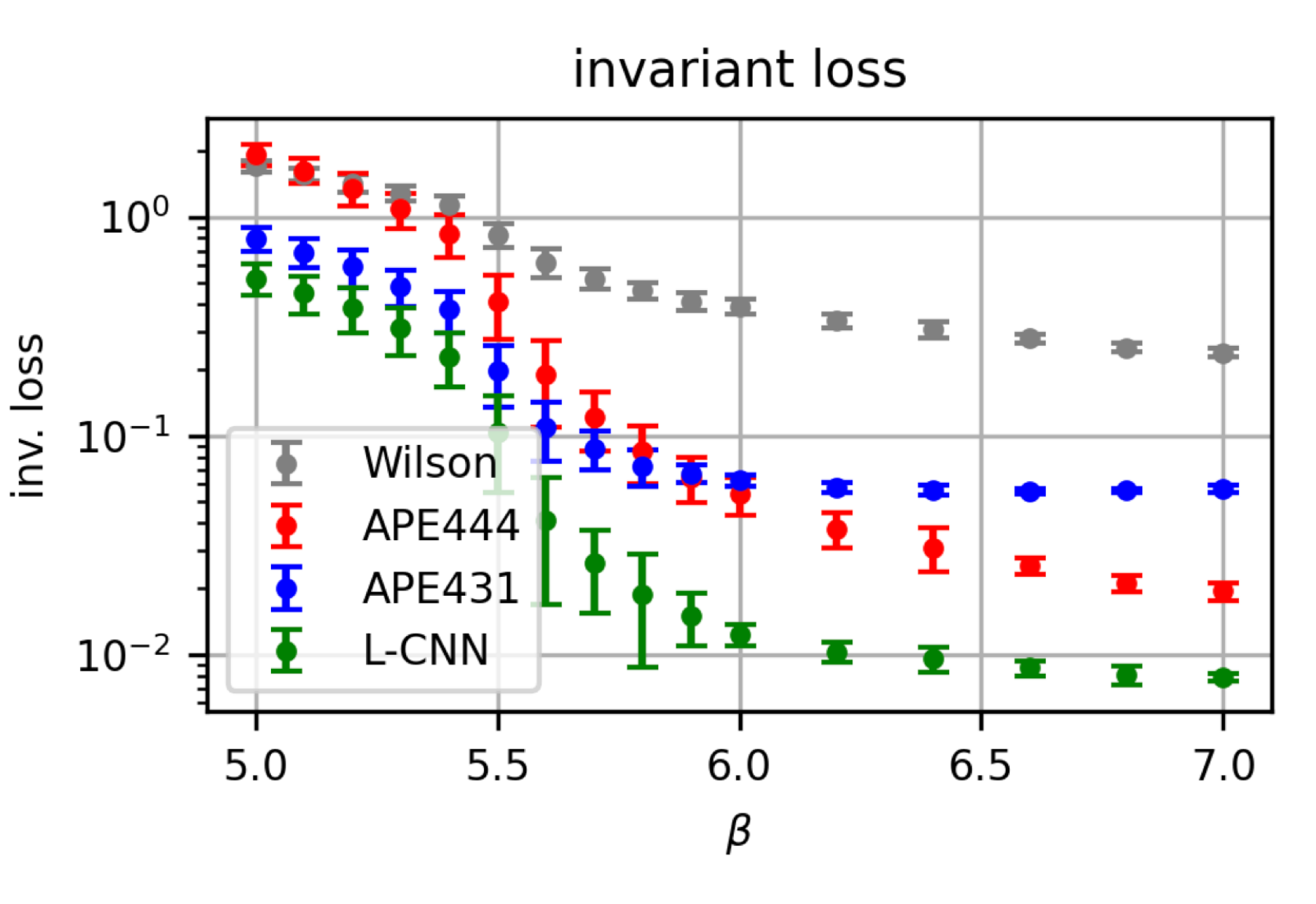}
   \includegraphics[width=0.45\textwidth]{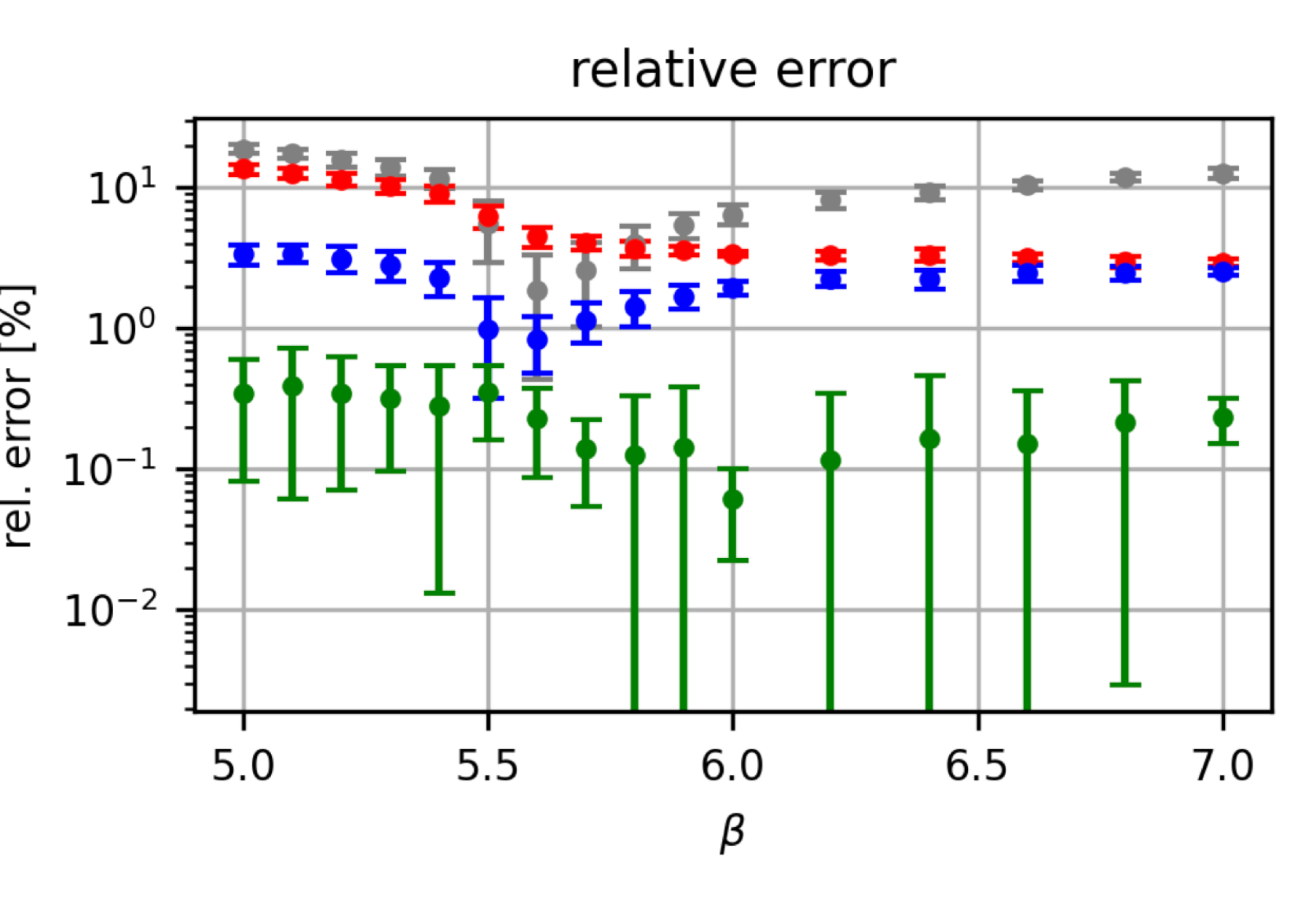}
   \includegraphics[width=0.9\textwidth]{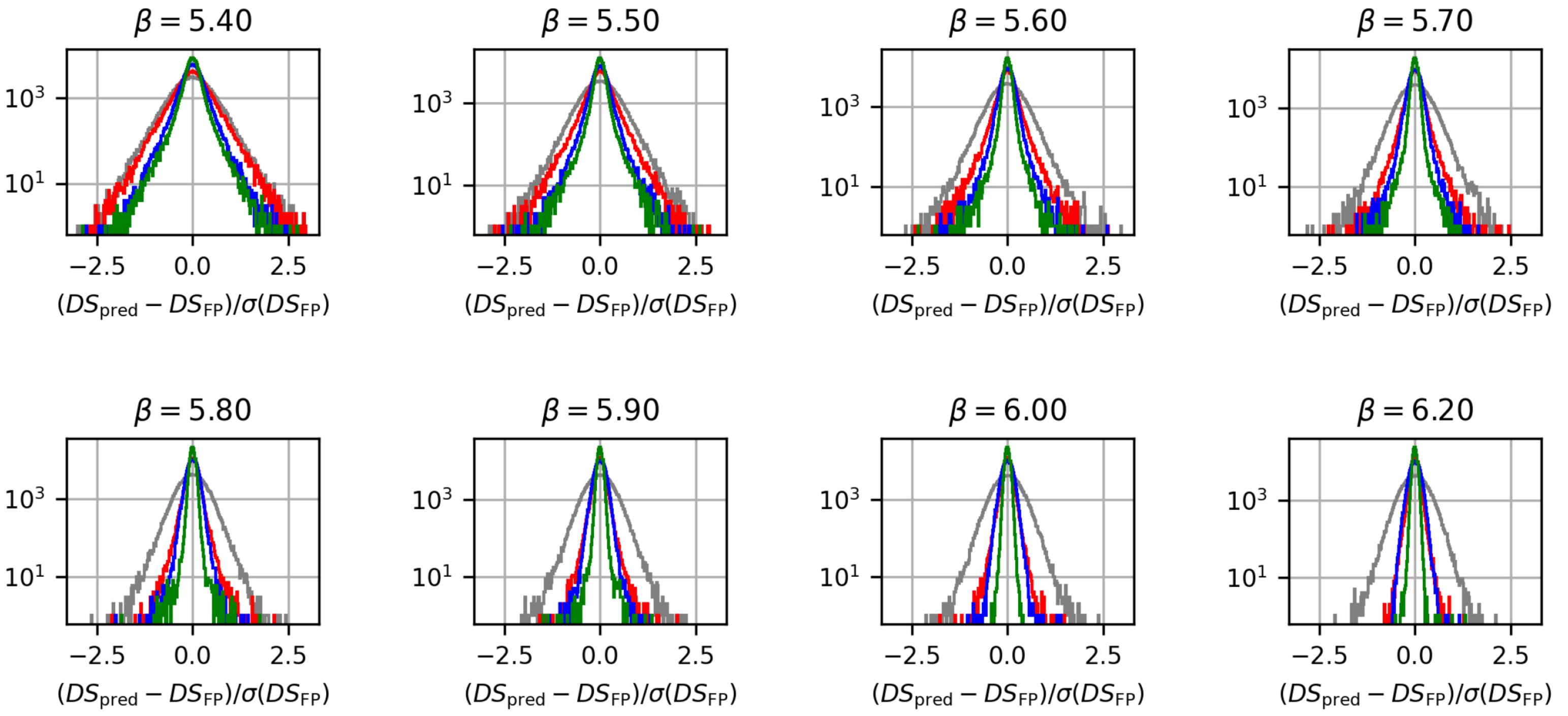}
  \caption{Results for the FP action parametrized by a particular
    L-CNN model as described in the text. We
    show the invariant loss ({\it top left plot}) and the relative error
    on the action values ({\it top right plot}) evaluated on the
     ensembles generated with various values
    of $\beta$ using the Wilson action. The plots in the lower two rows show the 
    distributions of the difference between true and predicted derivatives.
    For comparison, we also show the
    results for the Wilson gauge action and the APE444 and APE431
    parametrizations of the FP action. 
    \label{fig:results_derivatives}}
\end{figure}

In
addition, we make use of the derivatives of the FP action with respect
to the gauge links. They are determined through the FP equation and
are given by the derivatives of the blocking kernel,
\[
  D^\text{FP}_{x,\mu,a}[V] \equiv \frac{\delta A^\text{FP}[V]}{\delta V^a_{x,\mu}} = \frac{\delta
    T[U,V]} {\delta V^a_{x,\mu}} = -\kappa \, \text{Re Tr}(i t^a
  \,V_{x,\mu} Q^\dagger_{x,\mu}) \, ,
\]
i.e., they implicitly depend on the minimizing configuration
 through $Q^\dagger_{x,\mu} = Q^\dagger_{x,\mu}[U]$. The
derivatives yield $4 \times (N_c^2-1) \times L^4$ data per
configuration, one data point for each link and color index. Apart from providing an immensely larger amount of FP data for training the L-CNN, the derivatives are particularly
suitable in the ML procedure, because they are automatically accessible
through the backpropagation process as they are the derivatives of
part of the loss function w.r.t.~the input.
The ML loss function, which is minimized during training, is
defined as a weighted sum of the following two contributions: 
\begin{align*}
   \mathcal L_1 &= \frac{1}{ L^4}\frac{1}{N_\text{cfg}} \sum_i \left|  A^\text{FP}[V_i] -
  A^\text{L-CNN}[V_i]\right|, \\ 
  \mathcal L_2 &= \frac{1}{8 (N_c^2-1)L^4}
\frac{1}{N_\text{cfg}} \sum_{i,x,\mu, a}  (D^\text{FP}_{x,\mu,a}[V_i] -
  D^\text{L-CNN}_{x,\mu,a}[V_i])^2.
\end{align*}
The first expression measures the absolute error of the action density. The second expression measures the error of the derivatives in a gauge invariant way, which we term \emph{invariant loss}.
Here, $N_\text{cfg}$ is the number of configurations while
$A^\text{L-CNN}$ and $D^\text{L-CNN}_{x,\mu,a}$ are the action and derivative values predicted by the L-CNN.
The data set we use for the
supervised learning is produced by first generating
ensembles of coarse gauge field configurations for a large range of
fluctuations using the Wilson gauge action with the corresponding sets labeled by the gauge coupling $\beta_\text{wil}$. For each configuration $V_i$, we then find the
configuration $U_i$ minimizing the RHS of Eq.~(\ref{eq:FP equation}) and 
producing the training data.

\section{Results}
We are currently in the process of 
evaluating different
L-CNN models and learning strategies. In Fig.~\ref{fig:results_derivatives} we show the results for the FP action parametrized by a particular
L-CNN model with three L-Conv and L-Bilin layers containing 12 channels each, parallel
transport with $k=\pm 1$ in the first two L-Conv layers, and no transport in
the third L-Conv layer.
Neither activation layers nor a traditional CNN is used after the final Trace layer. As baselines for comparison, we use the Wilson gauge
action, and two older, but rather expressive parametrizations of the
FP action denoted by APE444 and
APE431. The latter has been optimized to satisfy the FP
equation  specifically on coarse lattices and has been extensively
used and tested in MC simulations \cite{Niedermayer:2000yx}. 
We find that the L-CNN describes the FP action values
better than the best old parametrizations over a large range of gauge
field fluctuations corresponding to
$5.0 \leq \beta_\text{wil} \leq 7.0$, and with a relative error by
about one order of magnitude smaller. We see a similar improvement in
the description of the FP action derivatives.

\begin{wrapfigure}{r}{0.5\textwidth} 
\centering
\includegraphics[width=0.4\textwidth]{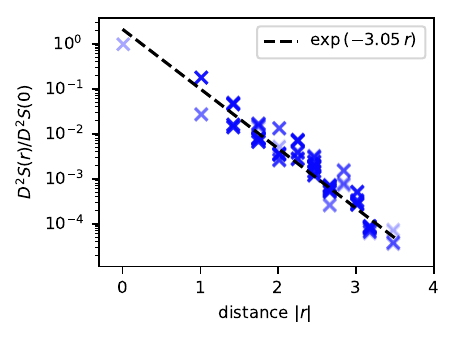}
\caption{\label{fig:locality} Measure of the gauge link couplings as a function of separation in lattice units for one specific L-CNN architecture including three layers.}
\end{wrapfigure}

A crucial point in the practical implementation of the FP program is how local the FP action is, i.e., whether the generated couplings are sufficiently short-ranged.  We probe the locality by calculating $\delta^2 A^{\text{L-CNN}}[V]/\delta V_{x,\mu}^a \delta V_{y,\nu}^b$ and forming a suitable, gauge-invariant norm $D^2S(x-y)$. The result 
on a coarse configuration at $\beta_\text{wil}=5.0$ is shown in Fig.~\ref{fig:locality}. We find that the couplings fall off exponentially, as desired, and take very small values at the distances where the L-CNN parametrization is truncated (by the choice of the number of layers and kernel sizes).

\section{Summary and conclusions}
Using highly improved gauge actions for generating gauge field ensembles holds the promise to overcome both the problems of {\it
  critical slowing down} and {\it topological freezing} when
approaching the continuum limit in a gauge field theory. 
This is
achieved by simulating the improved actions on coarse lattices, where
both problems are absent, while keeping the lattice artifacts under sufficient
control to allow a reliable and solid continuum limit. A
radical way to implement this approach is to use RGTs in order to
construct {\it quantum perfect actions} which have no lattice
artifacts at all. In a way, this program is similar in spirit to
those attempting to construct normalizing flows \cite{Kanwar:2020xzo,Boyda:2020hsi,Gerdes:2022eve,Bacchio:2022vje} or diffusion models \cite{Wang:2023exq, Wang:2023sry}
to overcome critical slowing down and topological freezing. In those
approaches, the invertible flows generate maps from a trivial or simple
distribution of gauge field configurations 
to a desired target distribution without including any physical
information apart from the target distribution. In contrast, the RG
approach makes use of the RGT flow in order to inform the map,
however, the RG flow is  of
course not invertible.

While quantum perfect actions have so far been elusive, {\it
  classically perfect FP actions} have been constructed and put to use
in the past.  As such they can immediately be employed in simulations
of four-dimensional SU($3$) gauge theories in order to overcome the
above-mentioned problems. In this work, we revisit the construction of
the FP actions and propose to make use of the latest developments in
designing L-CNNs and ML
techniques.
In this context, two crucial
questions arise. Firstly, can the FP action be parametrized
sufficiently well, or even better than before, with the new L-CNN
architectures? Secondly, is the FP action sufficiently local such that
any necessary truncation in the couplings is negligible?  Both
questions are addressed in these proceedings and answered in the
affirmative. In fact, it turns out that the L-CNNs are capable of
describing the FP actions to a higher accuracy than before and over a
much larger range of gauge field fluctuations.

The next task in our program is to investigate how well the L-CNN
parametrization of the FP action behaves in actual Monte Carlo
simulations and what its scaling properties are. The availability of
gauge-link derivatives of the FP action is the stepping stone for these further developments, since both
the HMC and the Langevin algorithms, as well as observables based on
the gradient flow, make use of derivatives.
The ultimate goal would of course be to apply exact
RGT steps. The results presented in these proceedings provide a
promising basis for further steps in that direction. 

{\bf Acknowledgments:} This work is supported by the US National Science Foundation under Grant No.~2014150, the Austrian Science Fund (FWF) projects P~32446, P~34455 and P~34764, and the AEC and ITP at the University of Bern.  The computational results presented have been achieved in part using the Vienna Scientific Cluster (VSC) and computing resources at the University of Bern.

\bibliographystyle{JHEP}
\bibliography{fpafcnn_PoS}

\end{document}